# Topological Anderson Insulator in cation-disordered $Cu_2ZnSnS_4$

Binayak Mukherjee*, Eleonora Isotta, Narges Ataollahi and Paolo Scardi*

Department of Civil, Environmental and Mechanical Engineering, University of Trento, Italy

* Authors to whom correspondence and material requests should be addressed: P.S. (paolo.scardi@unitn.it), B.M. (binayak.mukherjee@unitn.it)

## Abstract

Using *ab initio* calculations supported by experimental transport measurements, we present the first credible candidate for the realization of a disorder-induced Topological Anderson Insulator in a real material system. High energy reactive ball-milling produces a polymorph of $Cu_2ZnSnS_4$ with high cation disorder, which shows an inverted ordering of bands at the Brillouin zone center, in contrast to the ordered phase. Adiabatic continuity arguments establish that this disordered $Cu_2ZnSnS_4$ can be connected to the closely related $Cu_2ZnSnSe_4$, previously predicted to be a 3D topological insulator. Band structure calculations with a slab geometry reveal the presence of robust surface states, while impedance spectroscopy coupled with resistivity measurements point to the surface-dominated transport which such states would imply; thus making a strong case in favor of a novel topological phase. As such, this study opens up a window to understanding and potentially exploiting topological behavior in a rich class of easily-synthesized multinary, disordered compounds.

## Introduction

Topologically non-trivial materials present a novel and exciting field of research in condensed matter[1]. They are valued both for their importance to fundamental science as exotic states of quantum matter, as well as their inherent potential for application in new and future technologies including thermoelectrics[2–4], spintronics[5–7] and quantum computation[5,6,8]. Starting with the discovery of the Quantum Hall Effect (QHE) by von Klitzing et al[9], this class of materials has grown to include many candidates in 2-, 3-, and higher dimensional systems, a growing (albeit still small) fraction of which have been experimentally realized. 3D topological insulators (TI's) present a sub-class of these exotic materials and may generally be described as hosting insulating bands in the bulk with band-inversion at high-symmetry points, coupled with symmetry-protected gapless surface states[10]. In the absence of symmetry-breaking, these surface states support high-mobility electron transport along specific directions on the surface, without backscattering.

The first materials predicted and subsequently experimentally verified as 3D TI's were the alloy $Bi_xSb_{1-x}$[11] and the chalcogenides $Bi_2Se_3$, $Bi_2Te_3$, $Sb_2Te_3$[12–15], compounds where large spin-orbit coupling (SOC) was understood to be driving the topologically non-trivial behavior. Subsequently however, it was demonstrated by Fu[16] that topological surface states can also be protected by crystalline symmetries in the absence of SOC, leading to the so-called topological crystalline insulator (TCI's), and allowing for the possibility of topologically non-trivial materials with weak SOC[17]. In the meantime, the search for topologically non-trivial behavior was expanded from the binary compounds and their alloys to include ternary and quaternary materials. Among these are ternary half-Heuslers, ternary I-III-VI$_2$ and II-IV-V$_2$ chalcopyrites, I$_3$–V–VI$_4$ famatinites, and quaternary I$_2$–II–IV–VI$_4$ chalcogenides[18]. In particular, the structure of quaternary chalcogenides can be imagined as a (I–II)$_2$(II–VI)(IV–VI) sublattice with two zinc-blende formula units. It has been proposed[19] that the greater chemical and structural freedom in these compounds can lead to improved tunability of their band structures, thus making their study vis-à-vis topological properties worthwhile.

The possibility of TI's in the quaternary chalcogenide class was investigated by Chen et al[20], using density functional theory (DFT) band structures. They showed that HgTe, a 3D semimetal with the zinc-blende structure, may be transformed into a TI by introducing a strong crystal field splitting ($\Delta_{CF}$), either by epitaxial straining, or by substituting two group-II Hg ions with one group-I and one group-III ion, the latter approach resulting in ordered I-III-VI$_2$ chalcopyrites. Subsequently the authors found that, by replacing two group-III cations with one group-I and one group-II cation, thus forming I$_2$–II–IV–VI$_4$ chalcogenides, the non-trivial band gap of the materials could be increased further via a simultaneous increase in $\Delta_{CF}$ and the band-inversion strength (BIS). In a contemporaneous study, Wang et al[19] performed a DFT-based screening of several ternary famatinite and quaternary chalcogenides for TI's and were able to identify several naturally occurring, Cu-based 3D TI's. Unsurprisingly given weak SOC, Cu$_2$ZnSnS$_4$ (CZTS) was found to be topologically trivial, though the authors showed that it could be 'transformed' into a TI by changing the atomic number of the cations, manifested as a doping effect evolving towards the ternary TI Cu$_3$SbS$_4$.

Topological insulators, including the multinary compounds mentioned above, are generally known to host a bulk band gap coupled to gapless surface states, robust to weak levels of disorder. Several recent studies have highlighted how TI behavior can exist in aperiodic systems such as quasicrystals[21], and can persist in systems with bulk defects such as grain boundaries and vacancies below a certain threshold[22,23], nevertheless, sufficiently strong disorder is expected to close the bulk gap and destroy all topological features[10,23]. In light of this, a surprising prediction was made by Li et al[24], who claimed that adding disorder to otherwise trivial systems can lead to the emergence of topological behavior. Using tight-binding calculations, the authors showed that disorder-induced Anderson localization may lead to a renormalization of the topological mass of the charge carriers via the band structure, causing a transition from a topologically trivial phase to a TI, leading to the so-called Topological Anderson Insulators (TAI). TAI's have been theoretically shown to be feasible by introducing disorder into trivial 3D systems close to a topological phase[25,26]. TAI behavior was recently demonstrated by Meier et al[27] using quantum simulations in a metamaterial consisting of a 1D chain of ultracold rubidium atoms; nevertheless, evidence of TAI phases in real material compounds remains conspicuously absent.

Crucially, the quaternary chalcogenides screened in the aforementioned studies including CZTS are all ordered tetragonal structures. However, it is known though that CZTS crystallizes in multiple polymorphs. In a recent study, Isotta et al[28] demonstrated remarkably improved thermoelectric properties in a cubic polymorph of CZTS with complete cation disorder, synthesized using high energy reactive ball-milling, compared to the ordered tetragonal polymorph. Using DFT band structure calculations, we argue that introducing full cation disorder in CZTS drives it into a TAI phase; experimental measurements of resistivity and impedance spectroscopy are in agreement with the surface-dominated transport which such a phase is expected to host. As such, we present the first concrete prediction of a TAI in a material, opening up myriad possibilities for investigating topologically non-trivial behavior in disordered quaternary compounds.

**Band inversion in the bulk**

CZTS is a quaternary chalcogenide compound extensively investigated for its potential applications, primarily in photovoltaics[29–34], and recently thermoelectrics[28,35–40]. The most ubiquitous polymorph of CZTS is the kesterite structure (Figure 1a), which crystallizes in the tetragonal *I-4* space group. The structure may be described as alternating layers of cations and sulfurs, with a further alternation in the cation layers, which are either composed of Cu and Zn or Cu and Sn. Above 533K, this *I-4* structure undergoes the so-called order-disorder transition[41] into a tetragonal *I-42m* phase, the disorder being manifested through an in-plane randomization of the cations in the Cu-Zn layer (Figure 1b). The disorder induces a narrowing of the band gap compared to the ordered tetragonal polymorph (Fig 1d and 1e), while the increase in global symmetry introduces a 3-fold degeneracy at the center of the irreducible Brillouin zone (Γ-point) in the valence band maximum (Figure 1e). In a recent article, Isotta et al[28] have presented the synthesis of a novel polymorph of CZTS, this time with full cation disorder manifested as a complete randomization of atoms in the cation position (Figure 1c). This polymorph was found to crystallize in the cubic zinc-blende/sphalerite structure with space group *F-43m* and electronic structure calculations revealed the presence of significant inhomogenous bonding. This removes the 3-fold degeneracy present in the bands (figure 1f) of the disordered tetragonal polymorph, via strong crystal field splitting, while opening up the band gap somewhat (see Supplementary Note 1).

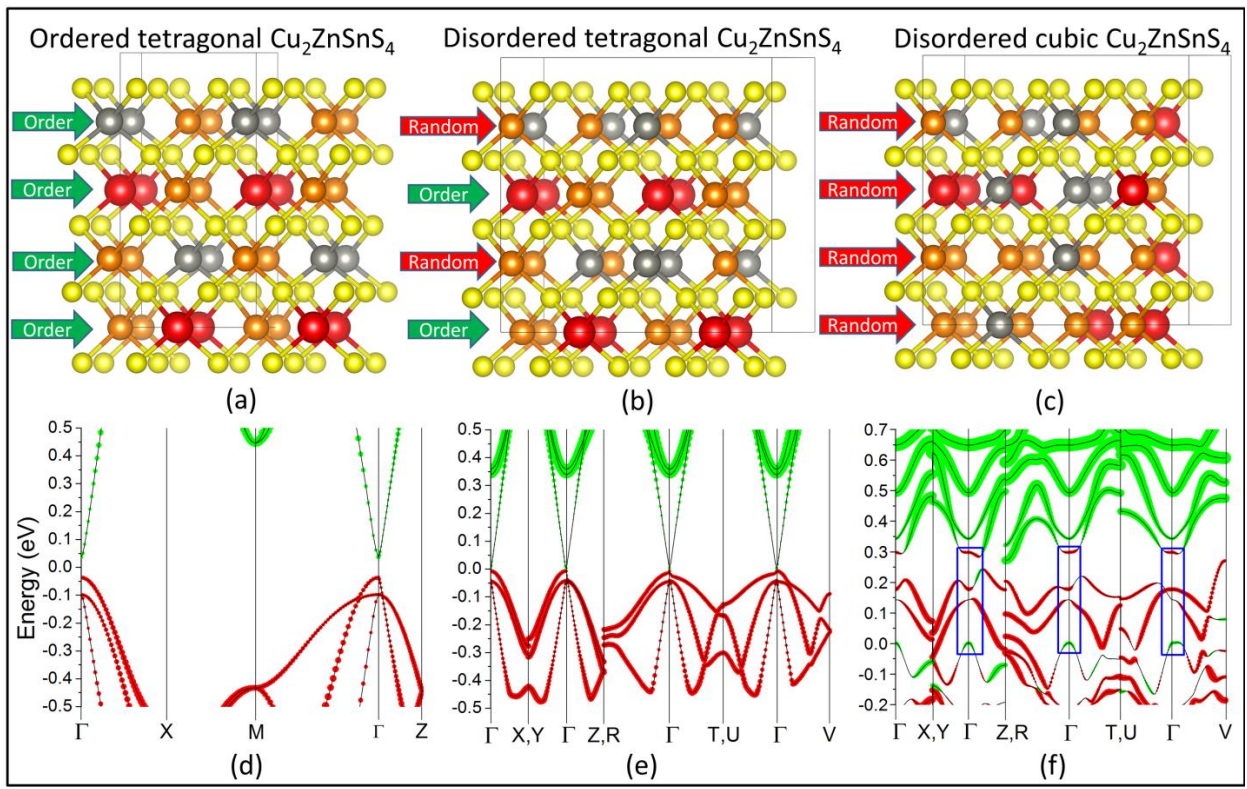

Figure 1. Crystal geometries and orbital-projected bands for the different polymorphs of CZTS: (a) shows the geometry of ordered tetragonal CZTS; (b) shows a supercell of disordered tetragonal CZTS; (c) shows a supercell of disordered cubic CZTS; Orange balls refer to copper ions, red balls to tin, gray balls to zinc and yellow balls to sulfur; green arrows identify the ordered layers, while red arrows show the layers with cation randomization; (d) shows the bands for ordered tetragonal CZTS, (e) the bands for disordered tetragonal CZTS and (f) the bands for cubic CZTS; green circles correspond to dominant contribution from sulfur-p orbitals, while red circles correspond to contribution from copper-d orbitals; the blue box highlights the region of band inversion in cubic CZTS.

A common feature of all three polymorphs of CZTS is that the states in the valence band maximum (VBM) are dominated by the Cu-d electrons, while those in the conduction band minimum (CBM) are mainly derived from S-p orbitals (figures 1d, 1e, 1f). However, upon closer inspection of the projected bands for cubic CZTS, we observe that the order of the bands is reversed at and around the Γ-point, with an inversion in the Cu-d and S-p orbitals (figure 1f). True disorder in ionic positions is of course rather difficult to simulate within the size-constraints of a DFT supercell with periodic boundary conditions, which impose a long-range order on the system; as such, in order to ensure that the band inversion is not an accidental artifact but rather a property of the system, we have calculated the band structure for a further 9 different configurations of cubic CZTS (supplementary Figure S1), with Cu, Zn and Sn ions randomly assigned to each cation position, while maintaining the $Cu_2ZnSnS_4$ stoichiometry (see Table T1 in supplementary information file) for energies of each configuration. The lowest energy configuration is shown in Fig 1c and 1f).

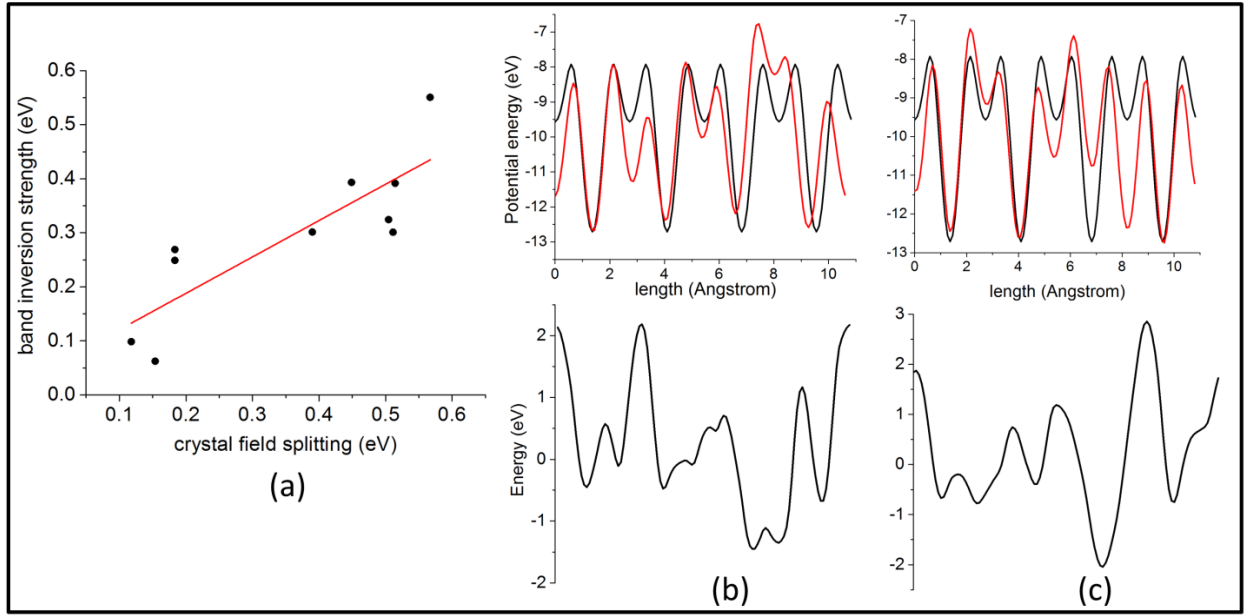

Figure 2. (a) shows correlation between band inversion strength and crystal field splitting; (b) shows the local potential in ordered tetragonal (black line) and cubic (red line) CZTS along the x-direction (above) and the difference between the two (below); (c) shows the local potential in ordered tetragonal (black line) and cubic (red line) CZTS along the y-direction (above) and the difference between the two (below).

The features of the bands are necessarily somewhat different from each other – each configuration of disorder generates a different kind of inhomogeneity in the charge distribution leading to different levels of crystal field splitting – crucially, however, band-inversion is present in every case, coupled in most cases with an anti-crossing (camel's back) feature at VBM and CBM. In fact, the band-inversion strength, as defined by the energy difference between the lowest inverted S-p level in the valence band and the highest Cu-d level, is found to be reasonably positively correlated with $\Delta_{CF}$, with a Pearson's r value of 0.84 (Figure 2a). While SOC is known to play a driving role in most topologically non-trivial systems, previous studies[42] have shown that it is negligible for tetragonal CZTS. We confirm that this remains the case in the cubic polymorph: including SOC in the calculation does not significantly alter the nature of the bands (supplementary figure S2b) at the valence and conduction band extrema, compared to the bands obtained without SOC (supplementary fig S2a), with band-inversion remaining intact. This points to the fact that SOC might not be the main feature driving the system into a TI phase.

Instead, we assert that it is the large $\Delta_{CF}$ which causes the inversion and opens up a non-trivial band gap, in line with the arguments proposed by Chen et al[20] in the case of strained HgTe and tetragonal ternary chalcopyrites and quaternary chalcogenides. Here, the $\Delta_{CF}$ is a result of inhomogeneous bonding, arising from the full cation disorder in cubic CZTS. In light of this disorder-induced topological transition, we propose cubic CZTS as a candidate Topological Anderson Insulator.

It is well known[43] that an inverted band structure corresponds to a negative (topological) effective fermion mass, $m$. In the TAI phase, Groth et al[25] have demonstrated that this inversion is obtained as a result of elastic scattering from a disorder potential, which leads to states with a definite momentum decaying exponentially as a function of space and time. When the effective Hamiltonian of the disordered system acts on the exponentially decaying state, it adds a negative correction $\delta m$ to the effective mass. This renormalized topological mass $m' = m + \delta m$ can have a sign which is opposite that of the bare mass $m$, corresponding to a band inversion. The low energy Hamiltonian $H$ of a general 3D topological Anderson insulator was written by Guo et al[26] as,

$$H = H_0 + \sum_j U_j \Psi_j^\dagger \Psi_j \qquad (1)$$

where $H_0$ is the Hamiltonian of the ordered (trivial) system, $\Psi_j$ is the overall wave function at the j-th lattice site, and $U_j$ is the on-site disorder or Anderson potential. By definition, the Anderson potential must vary randomly within the crystal lattice, and will correspond to a random component in the local potential energy in addition to the periodic component due to the crystal lattice.

Using DFT, we compute the local potential energy along the X-, Y- and Z-directions in the CZTS supercells, in order to compare the differences in the potential for the different polymorphs. It is evident from figures 2b and 2c (and supplementary figure S3a) that the potential in the ordered tetragonal polymorph (black curve) exhibits a highly periodic nature, deviating from this periodicity for the cubic (red curve) structures. We assert that the potential in the disordered polymorphs can be safely approximated to be the potential of the ordered structure plus a modifying potential due to disorder, in the spirit of equation 1. This modifying term is then given by the difference between the ordered and disordered potentials, as seen in figures 2b and 2c (and supplementary figure S3b).

Critically, it has been demonstrated[44] that bond-disorder, which adds random hopping terms to the Hamiltonian, and is present in many material systems, cannot drive a system into the TAI phase. As such, the bonding inhomogeneity prevalent in disordered CZTS cannot be held responsible for the non-trivial nature of the system, rather, it is an independent byproduct of the same random on-site cation disorder potential which also gives rise to the TAI behavior. This is evident from Fig 2b and 2c, which show the random disorder potential in the x- and y-directions respectively, which in fact are not the bonding directions in CZTS, thus putting our results in agreement with those of Song et al[44].

Girschik et al[45], instead, have suggested that any long range correlations in the disorder potential might lead to a strong suppression of the TAI phase. Such correlations can be reasonably precluded from our system by considering the global nature of the disorder in CZTS, constituting a total randomization of atomic species in the cation lattice sites. The nature of the disorder thus prevents long range correlations, instead promoting short ranged, random variations of the local potential, and allowing for the TAI phase to manifest.

It is then clear that the modifying potential is a highly random and aperiodic short-range onsite potential, making it a suitable candidate for the on-site disorder potential in the theory of TAI's. Given previous predictions of the closely related $Cu_2ZnSnSe_4$ as a TI[19], and the presence of a strong Anderson potential in the cubic polymorph, we put forward that ordered CZTS is driven into the cubic TAI phase by introducing a high level of cation disorder such as can be achieved through high energy ball-milling.

## Adiabatic continuity and topological surface states

While the presence of band-inversion in the bulk is considered a necessary condition for topological insulators, it does not on its own guarantee the presence of a topologically non-trivial phase[18]. To this end, adiabatic continuity arguments have emerged as a powerful tool to characterize the topological nature of materials through *ab initio* calculations, and have been used to predict new TI phases[46–50]. The process involves connecting a known topological material to a new structure through a series of adiabatic changes which include straining the crystalline lattice, tuning the strength of SOC, and modifying the nuclear charge of constituent atoms within the constraint of overall charge neutrality[18]. The argument is that if the Hamiltonian of this new system can be adiabatically connected to that of the known TI via some combination of the aforementioned ways, without inducing a band inversion or a closing of the gap, the new material can be considered to be topologically equivalent to the known material, and thus also a TI. Previous studies[19,20] have adiabatically connected quaternary chalcogenides to the known TI HgTe (strained), via both ternary famatinites and chalcopyrites. Of these compounds, the closest to our present case is the proposed TI[19] $Cu_2ZnSnSe_4$ (CZTSe) with an *I-42m* tetragonal stannite structure.

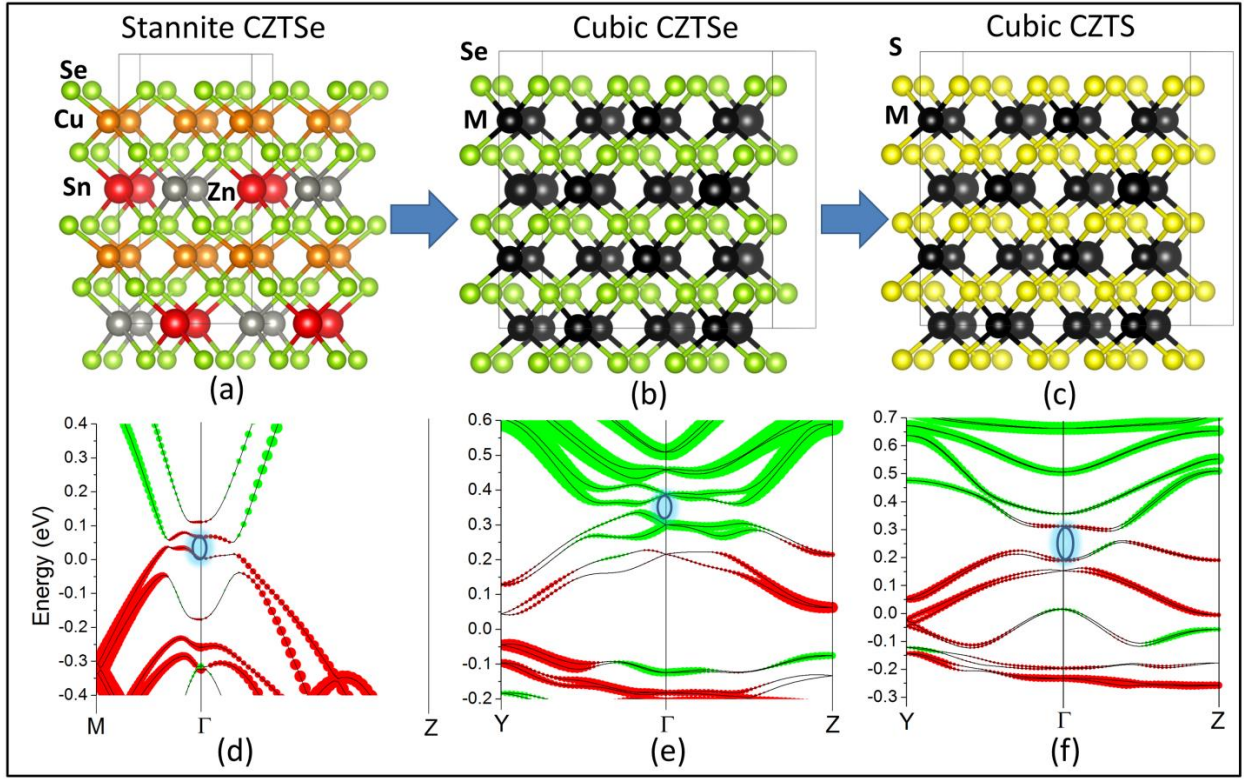

Figure 3. Adiabatic continuity between stannite CZTSe and cubic CZTS: (a) shows the crystal geometry of stannite CZTSe, (b) shows the geometry of cubic CZTSe, (c) shows the geometry of cubic CZTS); orange, red, gray and yellow circles have the same meaning as in Fig 1, green circles refer to selenium ions, black circles refer to a randomized cation; (d) shows the bands of stannite CZTSe, (e) the bands of cubic CZTSe, (f) the bands of cubic CZTS; red circles correspond to dominant contribution from Cu-d orbitals, green circles correspond to dominant contribution from anion-p orbitals; the black arrow shows the open non-trivial band gap.

Starting from this structure, we are able to transition to a fully disordered CZTSe with a cubic *F-43m* lattice by introducing randomization, interchanging the coordinates of a single pair of cations at a time. Given that both stoichiometry and charge remain conserved over-all, such a transition corresponds to an adiabatic change of the total Hamiltonian of the system (see table T2 in supplementary information file for energies of the intermediate configurations in the transition). Subsequently, we replace selenium ions with sulfur in the anion position thereby transitioning into our cubic CZTS. This, once more, is an adiabatic transition given that both S and Se are group IV elements with identical $s^2p^4$ outer shell electronic configurations, and the extra contribution of Se is only through fully occupied core levels that lie well away from the Fermi energy. As figure 3 demonstrates, this entire transition can be made without closing the inverted band gap at the Γ-point, thus allowing us to safely conclude that the disordered cubic polymorph of CZTS is in fact topologically connected to the previously predicted TI CZTSe in the adiabatic limit (see supplementary figure S4 for band structures of the entire transition). It should be noted that the high degree of cation disorder implicitly increases the global symmetry of CZTS from tetragonal to cubic with two interpenetrating sub-lattices of cations and sulfur anions, with the same *F-43m* space group as HgTe, the parent compound of this family of adiabatically connected 3D TI's.

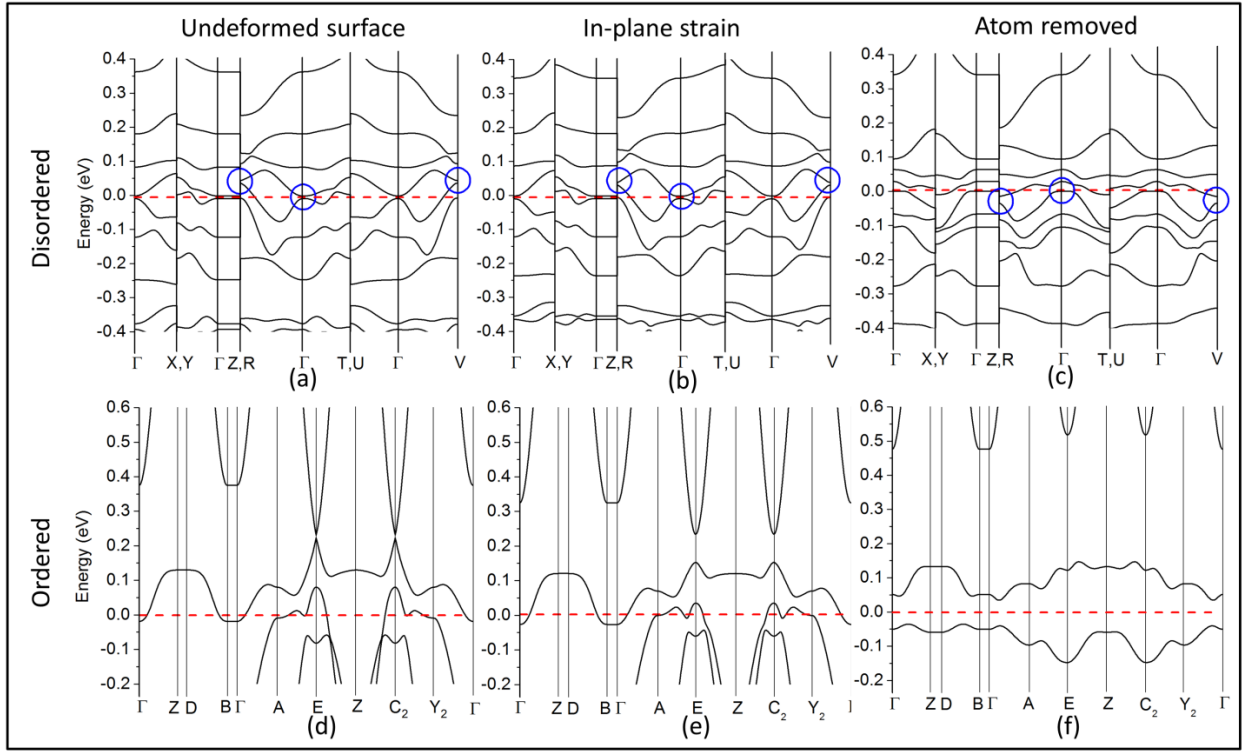

Figure 4. Surface states: (a) shows the surface band structure of cubic CZTS, (b) shows the corresponding surface bands under 2% expansive strain in the XY-plane, (c) shows the corresponding surface bands with one sulfur atom removed from the surface; blue circles mark the gapless surface states; (d) shows the surface band structure of ordered tetragonal CZTS, shows the corresponding surface bands under 1% expansive strain in the XY-plane, (c) shows the corresponding surface bands with one sulfur atom removed from the surface.

Further evidence to verify the topologically non-trivial nature of a material can be obtained by characterizing the topologically protected gapless surface states which are guaranteed through bulk-boundary correspondence. The calculation of these states from first principles is computationally demanding, and the results can be distorted by spurious gaps due to interactions between periodic copies[18]. In order to reduce these artefacts, we calculate the surface states for a 001 surface with sulfur termination within a slab geometry, with a large (10 Å) vacuum introduced in the Z-direction (see supplementary Fig S5 for simulation cell). Other than the expected quasi-gapless state at the Γ-point, our calculations point to a further two states at the high-symmetry R and V points (Fig 4a). This is in agreement with the requirement for an odd number of Dirac points, though it must be mentioned that these states tend more towards a parabolic curvature rather than the well-known linearly-dispersing Dirac cones. This "quadratic band touching" was predicted by Fu[16] using a tight-binding model for spinless fermions, which in real materials correspond to systems with weak SOC. These so-called "Schroedinger-paraboloids" have recently been reported in novel topological semi-metals, namely the Weyl semi-metal candidate $SrSi_2$[51] and the so-called Schroedinger semimetal $Be_2P_3N$[52], both of which, interestingly, show weak SOC. In light of these recent results, the non-linear dispersion of the bulk bands in disordered cubic CZTS is to be expected, and might be considered a general feature in the emerging class of topologically non-trivial materials with weak SOC.

Topological surface states can be sharply distinguished from well-known trivial surface states in semiconductors/insulators, insofar as the latter are less robust and can be removed via surface deformation[18]. Interestingly, we find that the topologically trivial ordered tetragonal CZTS hosts such surface states at the E and $C_2$ high symmetry points on the 001 surface (figure 4d). As expected, deforming the surface, for example by applying a small (1%) in-plane expansive strain (figure 4e), or simply removing a single S atom from the surface layer (figure 4f) leads to an opening of the gap, causing the trivial gapless states to vanish. In comparison, the surface states in the cubic polymorph are found to be significantly resilient (figure 4c, 4d) to identical surface treatments, confirming that these states are in fact topologically protected.

**Resistivity measurements and impedance spectroscopy**

Conclusive evidence in favor of a TI can be obtained via a direct experimental observation of the topological surface states with angle resolved photoemission spectroscopy (ARPES) – however, this technique is suitable only for large single crystals with perfectly clean surfaces. Disordered cubic CZTS, synthesized as it is via reactive ball-milling, exists essentially in a metastable, nano-polycrystalline form, and is therefore unavailable for ARPES. Under such circumstances, transport experiments can provide an alternate route to support the presence of the surface states: given a sufficiently insulating bulk, it is expected that conduction is dominated by the surface[53], a feature which can be verified through impedance spectroscopy[54,55] as well as by observing the grain size dependence of the conductance[56–59].

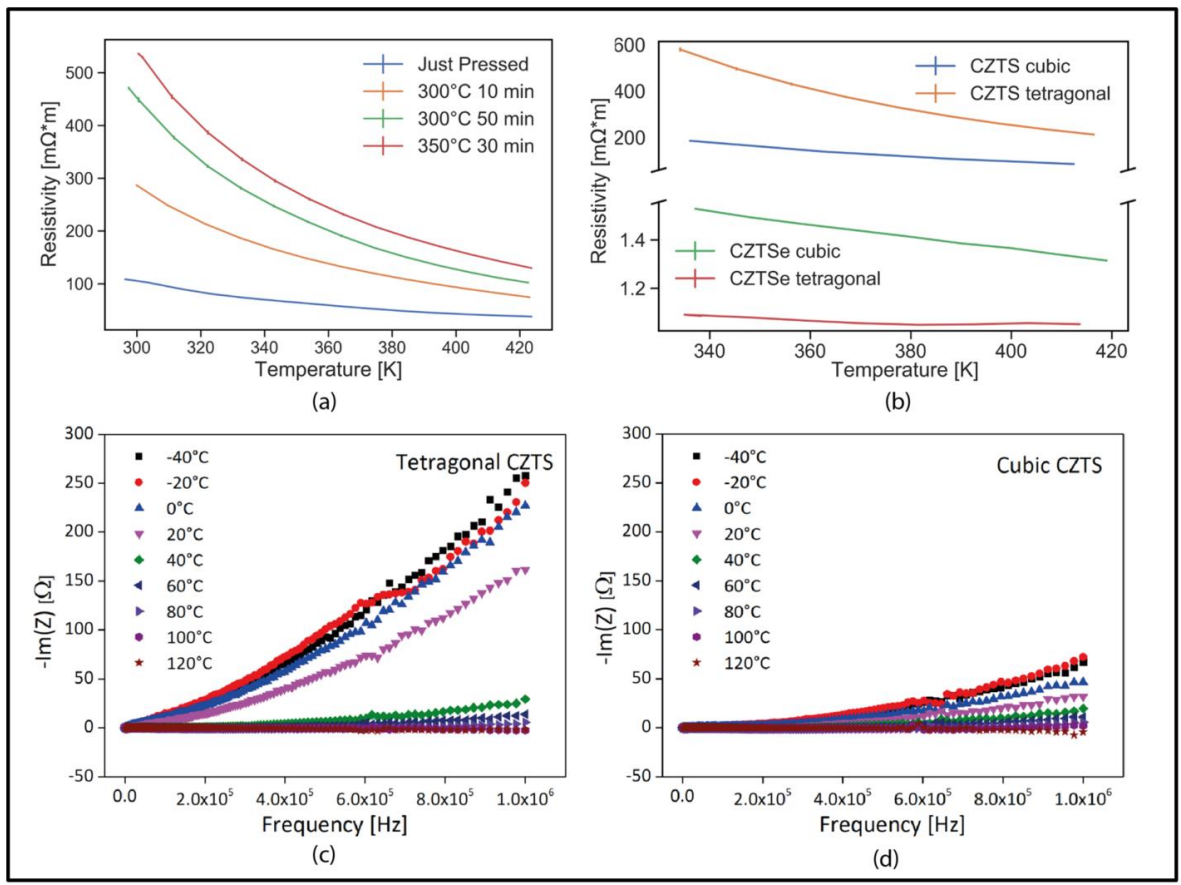

Figure 5. Transport measurements: (a) Electrical resistivity measurements performed for four different samples of cubic CZTS that undergone a different thermal treatment. The mean domain size is estimated from a Rietveld refinement of XRD with WPPM modelling and yields values of 13(1) nm, 26(2) nm, 29(3) nm and 36(3) nm, respectively with progressive thermal treatments. See supplementary material for details. (b) Comparison of electrical resistivity for cubic and tetragonal samples of CZTS and CZTSe. (c) Loss spectrum of disordered cubic CZTS at multiple temperatures from impedance spectroscopy. (d) Corresponding loss spectrum for ordered CZTS.

Indeed, experimental measurements on cubic CZTS samples seem to point to surface-dominated conduction. We have measured the grain size dependence of electrical resistivity for some cubic CZTS samples (Figure 5b). Each sample has been exposed to different thermal treatments, progressively longer and at a higher temperature, while keeping them below the cubic to tetragonal phase transition. Grain growth occurs with the progression of the thermal treatments, as confirmed by X-ray Diffraction (and Rietveld refinement with WPPM modelling[60], see supplementary figure S6). Concurrently, electrical resistivity increases (figure 5a and supplementary figure S7). This is in contradiction to a general trend in chalcogenide semiconductors: thermal treatments typically lead to higher density, improved connectivity between adjacent domains, and a reduction in carrier scattering from the grain boundary, all of which improving electrical conduction. The opposite seems to be the case for cubic CZTS, further supporting the possibility of surface-dominated conduction, compatible with a topological nature of the material[53]. Additionally, it should be noted that the difference in resistivity gets smaller with increasing temperature. This can be explained as an increasingly higher contribution from the grain (bulk) due to the thermal activation of carriers, while the surface contribution, arising from the possible topological surface states, should be scattering-resistant and therefore athermal.

To provide further experimental evidence, we synthesized a disordered cubic CZTSe sample with a ball-milling and sintering procedure equivalent to that of cubic CZTS (reported elsewhere[28]). Comparing the electrical resistivity of disordered cubic CZTS and CZTSe with the corresponding ordered tetragonal samples yields intriguing results. For CZTS, the disordered sample presents a considerably lower resistivity compared to the ordered, while the opposite is observed for CZTSe. Additionally, the two polymorphs of CZTS show a remarkable difference in resistivity of more than three times, while for the CZTSe phases only a minor difference of ~30% is observed. The atypical correlation between resistivity and sintering-level/grain size can be explained by the presence of topological surface states in the disordered polymorph, which are absent in the trivial ordered phase. For CZTSe, both the cubic (this work) and tetragonal[19] polymorphs are predicted as topologically non-trivial, and indeed a more typical trend of electrical resistivity is observed, with the more sintered sample presenting lower values of resistivity.

In addition to the resistivity measurements, we use impedance spectroscopy to measure the frequency dependence of the imaginary part of impedance (-Im(Z)), the so-called loss spectrum, to evaluate the transport behavior within the material. In tetragonal CZTS, the -Im(Z) value increases with the increasing frequency up to 20°C (Figure 5c, black, red blue and pink scatter plots), above which the variation of the –Im(Z) values is negligible (Figure 5c, green, indigo, purple, magenta and burgundy plots). In contrast, the corresponding measurement for disordered CZTS is seen to vary little over the whole range of temperature, pointing to a predominantly athermal transport mechanism. Additionally, the frequency at which -Im(Z) peaks can be extrapolated to be, are much higher for the disordered polymorph than the ordered. The peak frequency for surface-dominated transport is known to be much higher than for bulk, due to the larger resistance and capacitance of the bulk compared with that of the surface[61]. The higher frequency peak also corresponds to a faster overall movement of charge carriers[62–64], as one might expect in a system where transport is dominated by scattering-resistant topological surface states. Finally, the Nyquist plot from impedance spectroscopy measurements (supplementary figure S8) reveals a unique single semicircle, a feature typically connected with the response of the grain boundary[65,66]. This indicates that the surface component of conduction outweighs the bulk component.

While the provided experimental evidence falls somewhat short of being conclusive proof, it is nevertheless definitely in agreement with the results of the DFT calculations, suggesting that disordered cubic CZTS is in fact a topological Anderson insulator.

## Conclusion

In the present article, we propose a possible candidate for a disorder-induced TI material, the so-called Topological Anderson Insulator. High-energy reactive ball milling has recently been used to produce a low-temperature cubic ($F\text{-}43m$) phase of the quaternary chalcogenide $Cu_2ZnSnS_4$, with complete randomization in the cation positions. DFT calculations show that this novel disordered polymorph has an inverted band order in the conduction and valence band extrema at and close to the Brillouin zone center, in contrast to the trivial bands of the ordered tetragonal ($I\text{-}4$) counterpart. Furthermore, the band-structure of this phase is connected adiabatically to the bands for ordered tetragonal ($I\text{-}42m$) $Cu_2ZnSnSe_4$, known to be a 3D TI, without closing the inverted band-gap. Surface slab calculations reveal the presence of an odd number (three) of quasi-gapless surface states, which are remarkably robust to surface deformation such as strain and defects, sharply in contrast to fragile surface states in ordered CTZS. The resistivity, measured in multiple samples of disordered CZTS with different levels of sintering, show an atypical positive correlation with sample density and grain size – smaller grains with a higher surface-to-volume ratio appear to be conducting better. Measurements of the loss spectrum and Nyquist plot, obtained by impedance spectroscopy, further point to athermal, surface dominated transport in the disordered polymorph, in contrast to the ordered. These experimental features are in general agreement with the presence of gapless surface states, possibly promoting scattering-resistant transport. This serves to further explain the remarkable improvement in thermoelectric properties of disordered CZTS reported previously, primarily through an increase in conductivity without a corresponding tradeoff in the Seebeck coefficient, via surface dominated conduction.

In general, the evidence presented above, while strongly suggesting that disordered cubic CZTS behaves as a TAI, cannot be considered to be conclusive. Further investigation, both theoretical and experimental is part of ongoing research. Theoretical calculations using tight-binding models and effective Hamiltonians can be used to calculate the Berry phase and topological invariant, in order to provide a more fundamental understanding of the topologically non-trivial behavior of the material. On the experimental side, while the nano-polycrystalline nature of the material makes ARPES unlikely, further transport measurements at ultra-low temperatures, and in the presence of magnetic fields might be used to better characterize the nature of the surface states.

Nevertheless, our results indicate a strong possibility for the presence of novel topological phases in the realm of multinary, disordered compounds, where topological behavior is only partially understood. Such materials, easily and cheaply synthesized in comparison to perfect crystals for traditional TI's, open up diverse possibilities not just for fundamental research, but also the application of topological properties, particularly in the area of thermoelectrics.

## Methods

The *ab initio* electronic structure calculations have been performed using the plane wave basis set implemented in the Vienna ab initio simulation package (VASP)[67,68]. The electron-exchange correlation functional was approximated using the Perdew–Burke–Ernzerhof (PBE)[69] form of the generalized gradient approximation (GGA). It should be noted that the GGA tends to underestimate the band gap for most compounds, which may be corrected using computationally expensive hybrid functionals. However, a hybrid Hartree-Fock/DFT study[70] has established that the band topology for both computational schemes (hybrid and PBE) are very similar, with the hybrid functional shifting the conduction bands to a higher energy. All calculations were performed with an energy cutoff of 300 eV and Gaussian charge smearing in the order of 0.01 eV. Calculations for CZTS were performed both with and without spin-orbit coupling, while calculations for CZTSe were made only with SOC. The geometry was optimized with a 2 × 2 × 2 Monkhorst Pack (MP) Γ-centered k-mesh for 64 atom supercells until Hellman-Feynman forces on each atom were converged to below 0.01 eV/ Å. SCF calculations were made with a similar 4 x 4 x 4 k-mesh, with electronic degrees of freedom relaxed until the change in the total free energy and energy eigenvalues were both smaller than $10^{-6}$ eV. Calculations for surface states were performed within a surface slab geometry, with a 10 Å vacuum layer in the Z-direction to minimize the interaction between periodic copies. Only the top 3 layers of the slab were allowed to relax, with lower layers held fixed. Geometry optimization for the surface slab was made with 2 x 2 x 1 MP k-mesh, while SCF calculations used a 4 x 4 x 2 mesh. For band structure calculations, the high-symmetry path in the Brillouin zone was obtained using SeekPath[71]. VESTA[72] was used for visualizing atomic geometries.

Details for sample preparation have been described elsewhere[28]. Rietveld refinement with WPPM was performed using TOPAS[60,73]. Electrical resistivity measurements are performed on disk shaped samples (diameter ~16 mm and thickness ~1.5 mm) in four-contact configuration with a Linseis LSR-3 machine. Measurements are performed under a static He atmosphere in the range from room temperature to 420 K with a heating rate of 10 K/min. Three measurements per temperature are performed to calculate the mean and standard deviation, which are used in the plots as value and error bar. In addition, a declared instrumental accuracy of 7% should be considered.

The impedance spectroscopy measurements were performed on a tetragonal and cubic sample using a BioLogic science instrument (SP-150). The sample was polished and sandwiched between two gold electrodes in a CESH sample holder with a frequency range of 100 Hz to 1 MHz and a fixed voltage of 0.01 V at the temperature range of -40°C to 120°C using an intermediate temperature system (ITS).

## Data availability

The data that support the findings of this study are available from the corresponding authors upon reasonable request.

## Acknowledgements

The computational time was provided by CINECA - Italian Supercomputing Facility, with the Project CZTS - HP10CONX70 and the Modelling and Simulation Project of University of Trento.

## Author contributions

B.M. was primarily responsible for making the DFT calculations, theoretical analysis, and writing and editing the manuscript. E.I. was primarily responsible for supplying the samples and making the resistivity measurements. N.A. was primarily responsible for making the impedance spectroscopy measurements and interpreting the results. P.S. was primarily responsible for supervising the research and participated in analyzing theoretical and experimental results and editing the manuscript.

## Competing interests

The authors declare no conflict of interests.

**Supplementary Information**

**Supplementary note 1:**

Despite appearances to the contrary in Fig 1(f), disordered CZTS does in fact have a global band gap, albeit a complicated one. In our previous article (Ref 25), we have experimentally measured the optical gap, which shows a large value of ~1.5 eV; notably however, the Tauc plot obtained from UV-vis spectroscopy shows large Urbach tailing, as seen in the image below (left), Fig S3 from Ref 25 (Isotta et al, Phys. Rev. Applied 14, 064073).

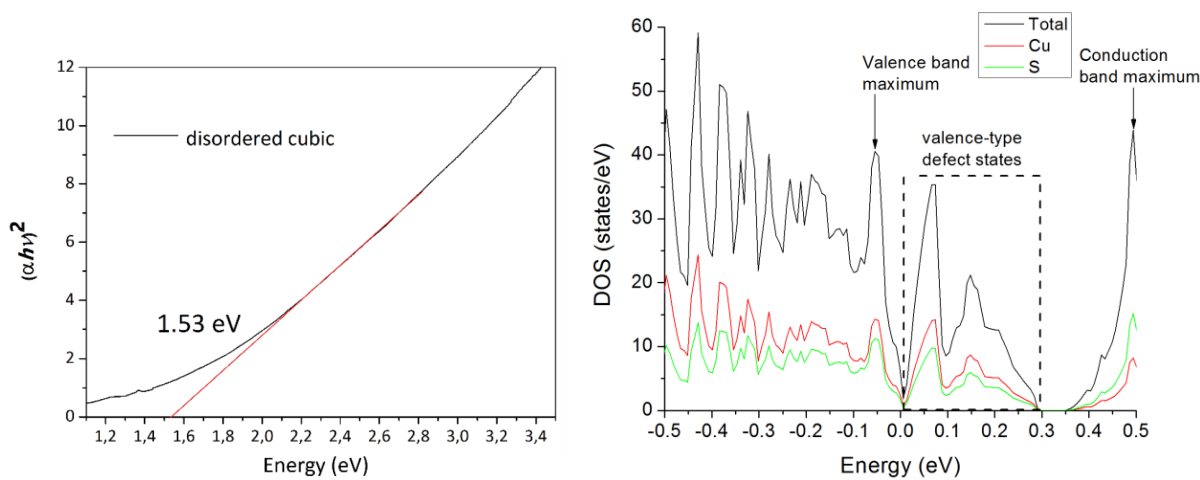

The calculated density of states (above, right) shows that the valence band is dominated by Cu-d electrons, while the conduction band is predominantly composed of S-p electrons, with the presence of large valence type defect states just above the Fermi energy, which is set to 0. The large Urbach tailing in the Tauc plots can be accounted for by these defect states, which of course arise due to the highly disordered nature of the material. For a detailed theoretical discussion on these defect states and how they improve experimentally measured thermoelectric properties, we refer the reader to Ref 28. These defect states, which originate from the same orbitals as the valence band, but lie mostly above the Fermi energy, also represent the valence band maximum seen in Fig 1f, with a smaller gap to the bottom of conduction band, allowing for band inversion. The fact that the Fermi energy lies within the valence band of course is characteristic of p-type semiconductors, and does *not* imply that the system has no global gap, but rather that it has band tailing and a reduced gap.

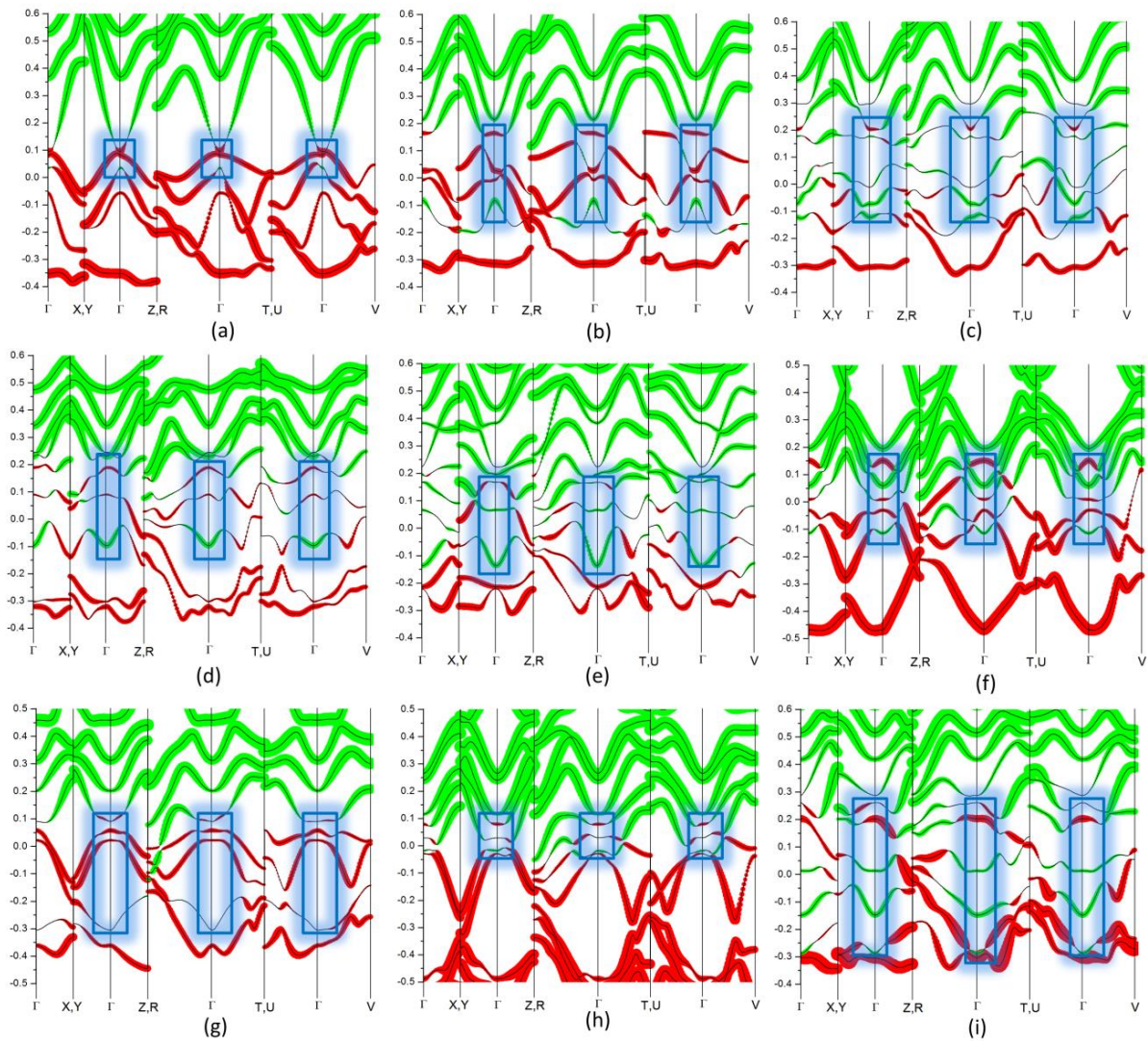

Figure S1. Bands for multiple configurations of disordered cubic CZTS. Symbols have save same meaning as in the main text.

| Configuration | Total energy |
|---|---|
| Config1 (Fig 1f) | -.25723222E+03 |
| Config2 (Fig S1a) | -.25487437E+03 |
| Config3 (Fig S1b) | -.25700924E+03 |
| Config4 (Fig S1c) | -.25462498E+03 |
| Config5 (Fig S1d) | -.25416696E+03 |
| Config6 (Fig S1e) | -.25364071E+03 |
| Config7 (Fig S1f) | -.25547144E+03 |
| Config8 (Fig S1g) | -.25522658E+03 |
| Config9 (Fig S1h) | -.25509927E+03 |
| Config10 (Fig S1i) | -.25400458E+03 |

Table T1. Total ground state energies of the multiple configurations of disordered cubic CZTS

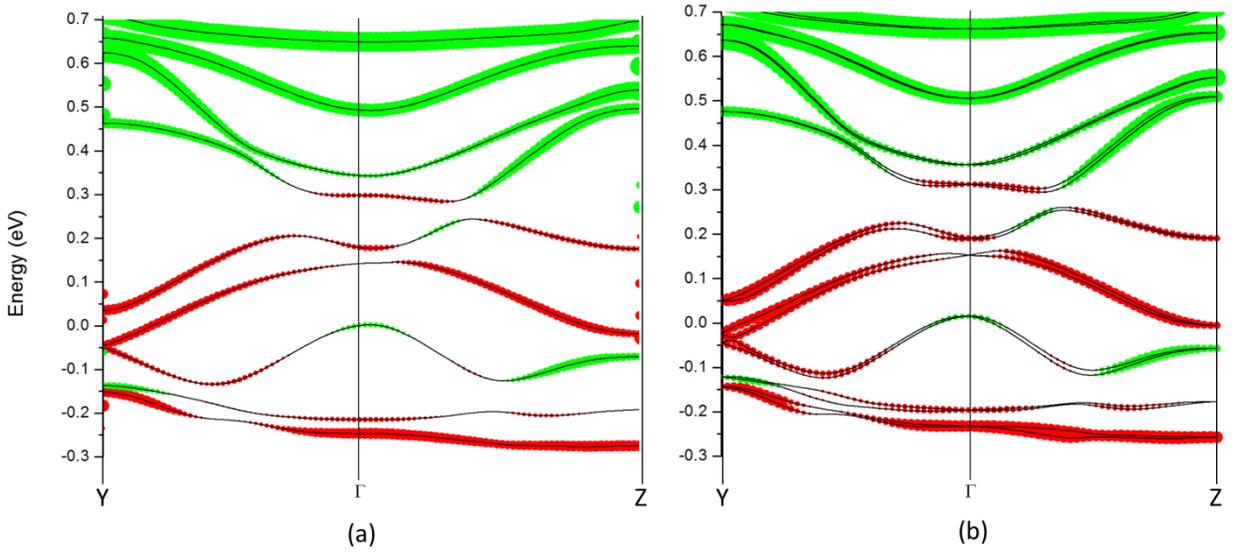

Figure S2. Comparison of bands for disordered cubic CZTS (a) without and (b) with spin-orbit coupling. Symbols have save same meaning as in the main text.

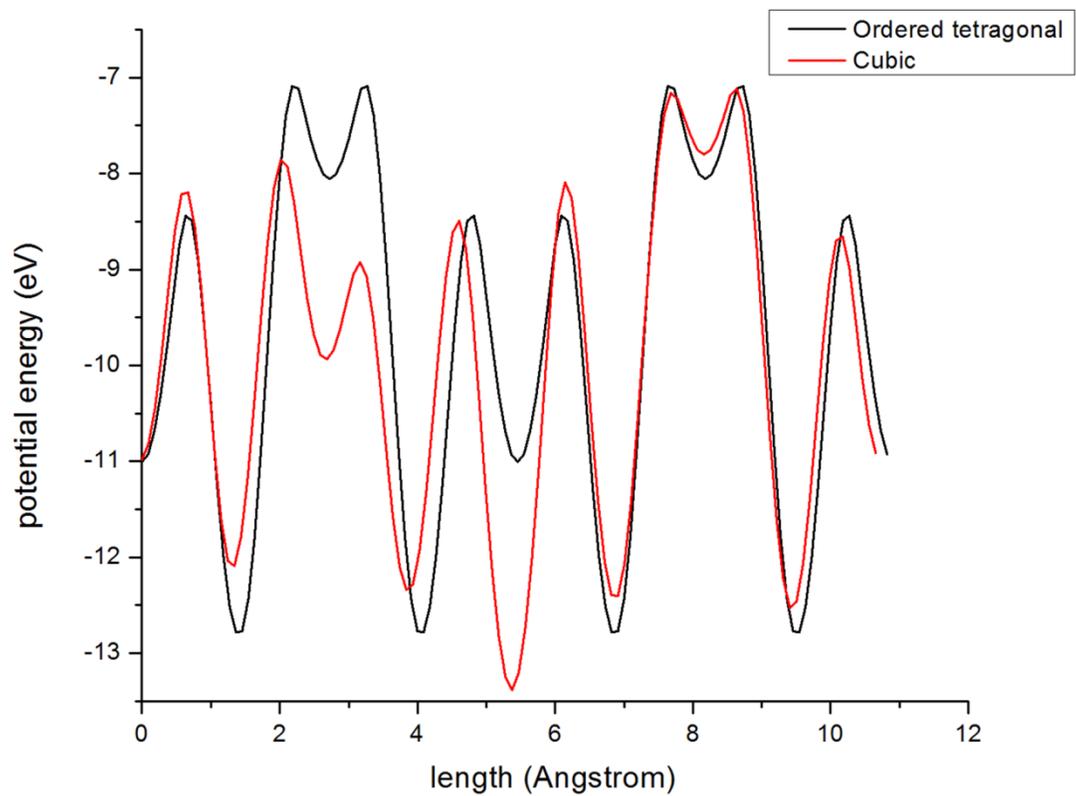

(a)

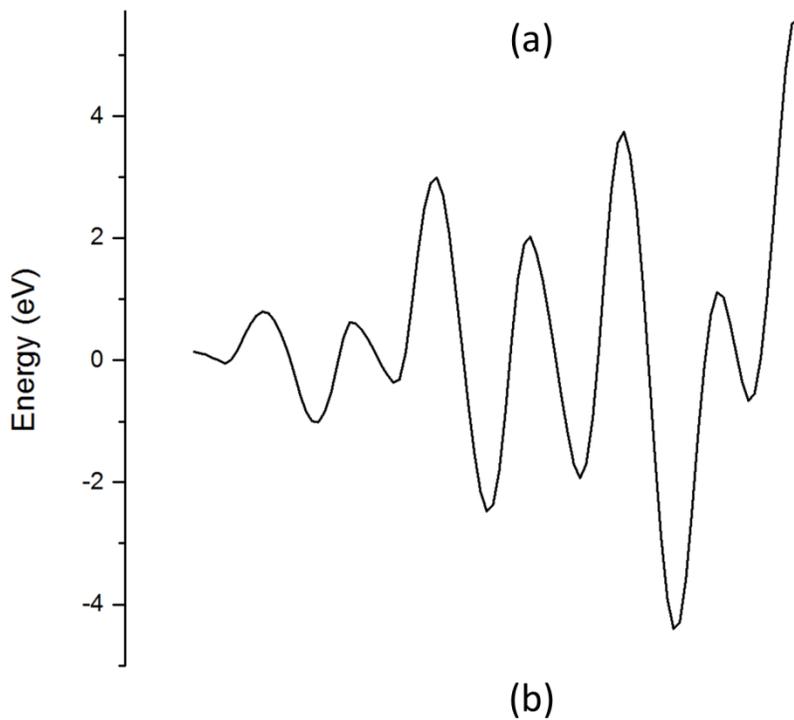

(b)

Figure S3. (a) Local potential in the Z-direction for ordered tetragonal (black) and disordered cubic (red) CZTS; (b) Difference between the potentials. The lattice mismatch in the Z-direction between the tetragonal and cubic supercells exaggerates the difference.

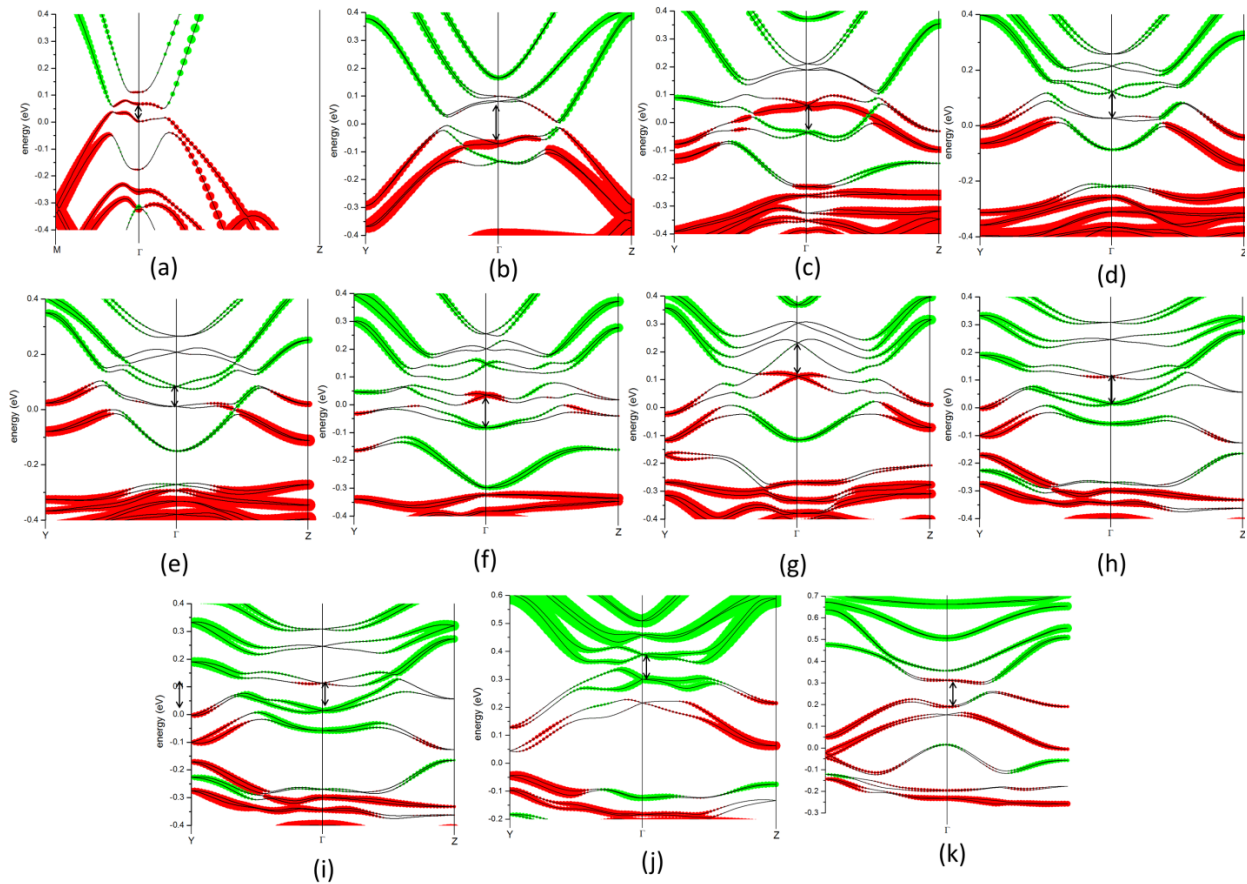

Figure S4. Adiabatic transition from (a) stannite CZTSe through (k) cubic CZTS. Intermediate images show the bands for intermediate states. Arrows highlight the open inverted band gap.

| Step | Total ground state energy |
|---|---|
| Step 1 (Fig S4a) | -.23556786E+03 |
| Step 2 (Fig S4b) | -.23552479E+03 |
| Step 3 (Fig S4c) | -.23484019E+03 |
| Step 4 (Fig S4d) | -.23303846E+03 |
| Step 5 (Fig S4e) | -.23233460E+03 |
| Step 6 (Fig S4f) | -.23221546E+03 |
| Step 7 (Fig S4g) | -.23220398E+03 |
| Step 8 (Fig S4h) | -.23231033E+03 |
| Step 9 (Fig S4i) | -.23122961E+03 |
| Step 10 (Fig S4j) | -.23132112E+03 |

Table T2. Total ground state energies for the intermediate configurations in the adiabatic transition from ordered to disordered CZTSe

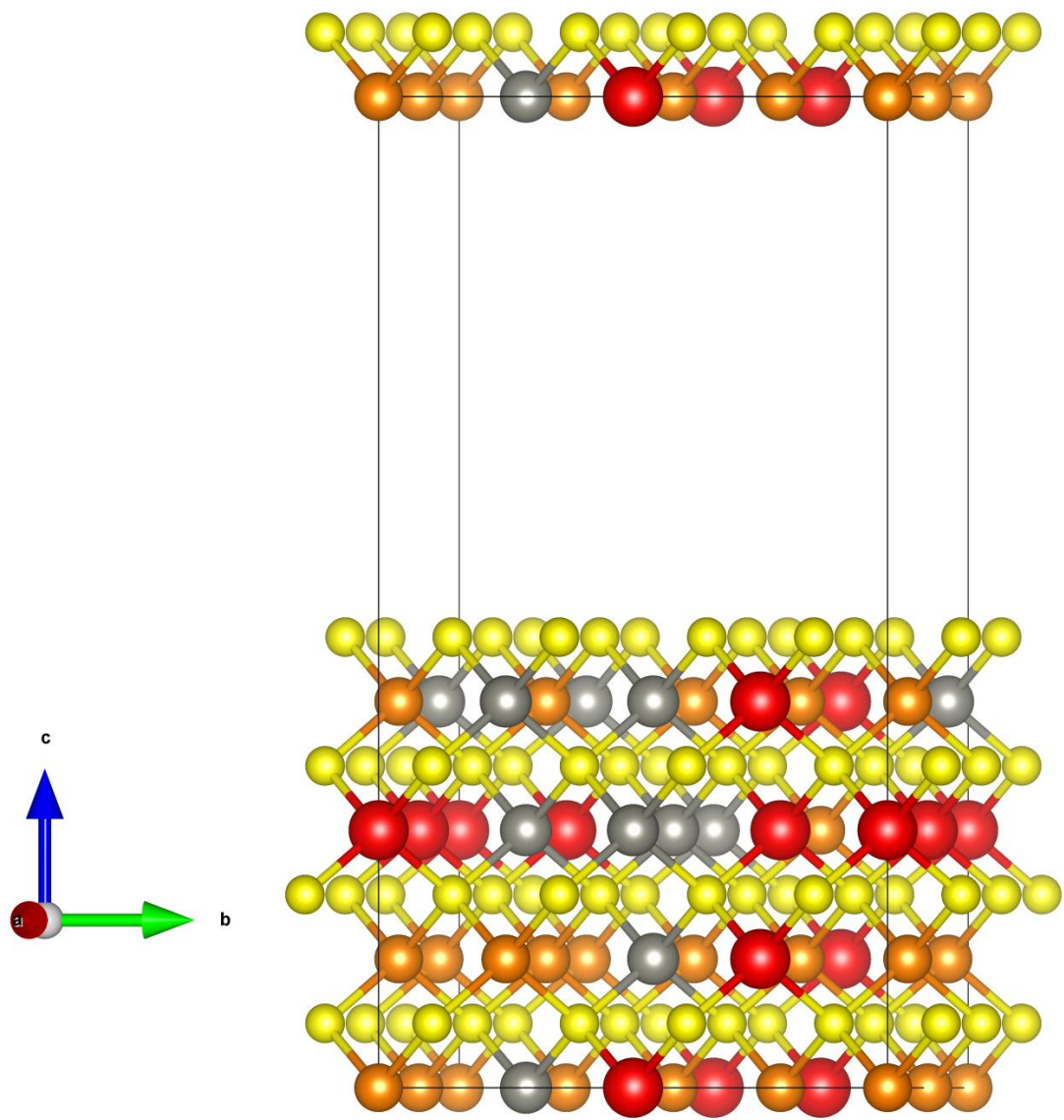

Figure S5. Surface slab geometry for the 001 surface of cubic CZTS with S termination. Orange atoms are Cu, grey atoms are Zn, red atoms are Sn and yellow atoms are S.

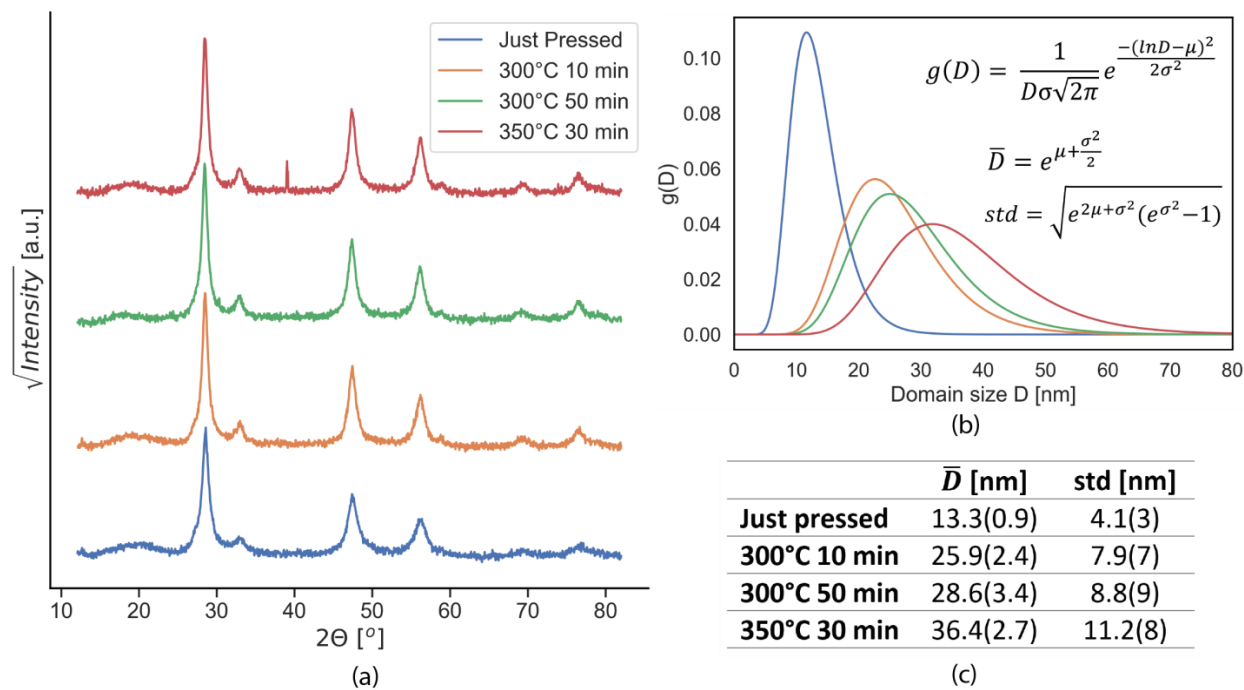

Figure S6. (a) XRD patterns for four different samples of cubic CZTS that undergone a different thermal treatment. The purpose was to increase the domain size while keeping the phase cubic. Rietveld refinement with the software TOPAS was performed on XRD with WPPM model for size distribution. This assumes a lognormal distribution of domain sizes, reported in panel (b) for the samples. Equations from top to bottom represent a lognormal distribution, the mean domain size and the corresponding standard deviation. It is visible how an increasing thermal treatment progressively leads to broader and right-shifted distributions, indicating a progressive grain growth. Distribution parameters are visible in Table (c).

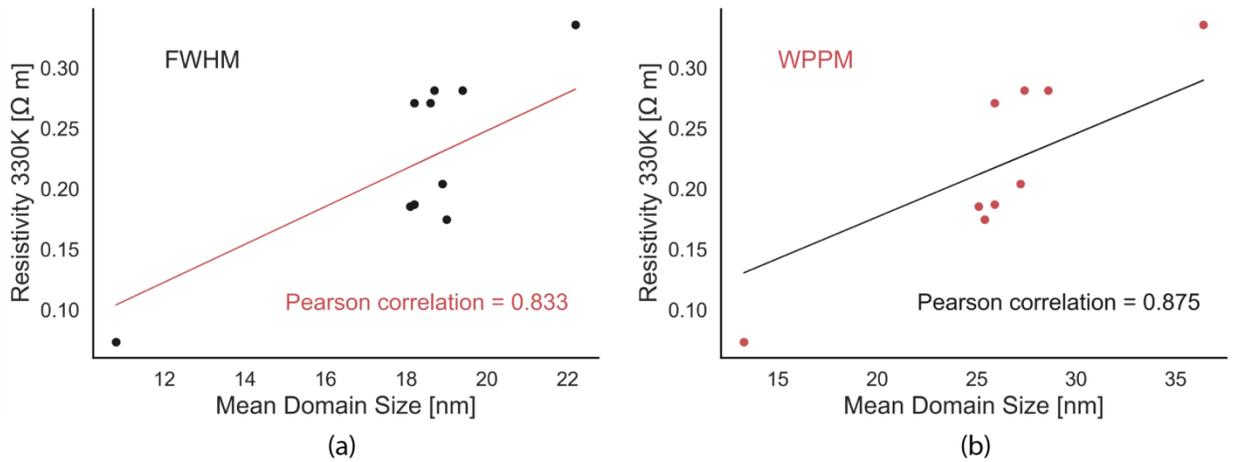

Figure S7. Electrical resistivity at 330K versus mean domain size for a collection of cubic CZTS samples that undergone different thermal treatments to achieve differential grain growth. The mean domain size was estimated from the Full-Width at Half-Maximum (FWHM) (a) and with the WPPM size macro (b).

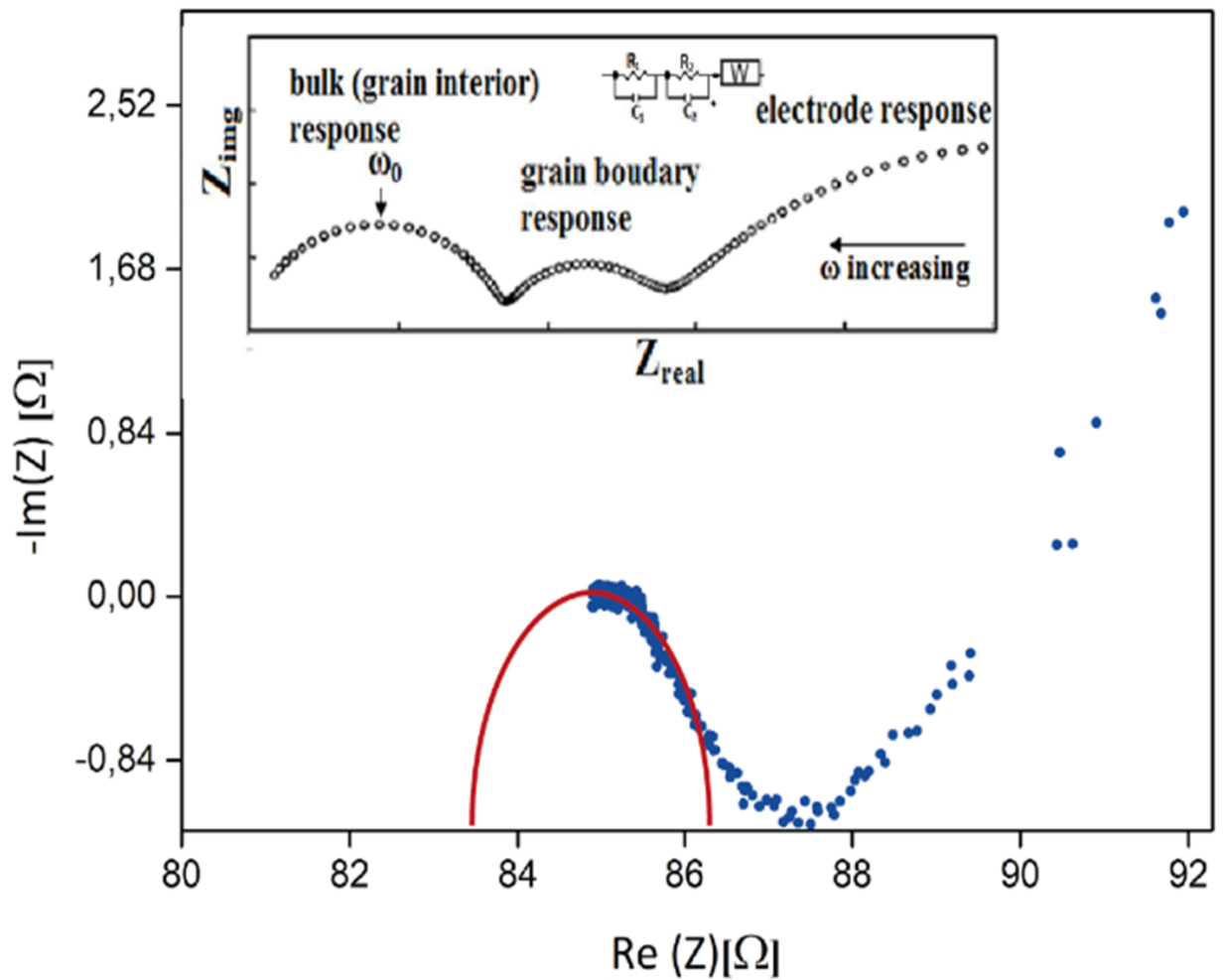

Figure S8. Nyquist plot from impedance spectroscopy measurements for a cubic CZTS sample at 278K with fitting (solid red). Inset in panel shows the typical Nyquist plot for grain, grain boundary and electrode response and the equivalent circuit. Commonly two semicircles are observed at higher and lower frequencies, corresponding respectively to the response of grain and grain boundary. When only one is present, this is connected with the response of the grain boundary.